\documentclass[amsmath,aps,prd,amssymb, tightenlines,nofootinbib]{revtex4-2}
\usepackage{graphicx}

\usepackage[utf8]{inputenc}
\usepackage{float}
\usepackage{graphics,graphicx}
\usepackage{dcolumn}
\usepackage{bm}
\usepackage{amsmath}
\usepackage{tensor}
\usepackage{slashed}
\usepackage{amsmath,amssymb,amsfonts,mathrsfs,bbold}
\usepackage{yfonts}
\usepackage{xcolor}
\usepackage{mathtools}
\usepackage{hyperref} 
\usepackage[framemethod=TikZ]{mdframed}
\usepackage{color}

\usepackage{multirow}
\usepackage{mathbbol}
\usepackage{mathtools}

\usepackage{slashed}
\usepackage{simpler-wick}

\usepackage[normalem]{ulem}

\usepackage{stackrel}
\usepackage[most]{tcolorbox}

\newcommand*{\Authoremail}{borden.75@buckeyemail.osu.edu}

\newcommand{\as}{\alpha_s}

\newcommand{\xoz}{x_{10}}
\newcommand{\xto}{x_{21}}
\newcommand{\xtt}{x_{32}}
\newcommand{\wint}{\int \frac{\mathrm{d}\omega}{2\pi i}}
\newcommand{\gint}{\int \frac{\mathrm{d}\gamma}{2\pi i}}

\newcommand{\dw}{\delta_{\omega}}
\newcommand{\gtwg}{G_{2\omega\gamma}}

\newcommand{\bas}{{\bar\alpha}_s}
\newcommand{\un}[1]{\underline{#1}}

\newcommand\pubdate{July 15, 2023}

\def\Title#1{\begin{center} {\Large #1 } \end{center}}
\def\Author#1{\begin{center}{ \sc #1} \end{center}}
\def\Address#1{\begin{center}{ \it #1} \end{center}}

\newcommand\pubblock{\rightline{\begin{tabular}{l}  \\ % Author's note number [if you need to add one] goes here
         \pubdate  \end{tabular}}}
\newenvironment{Abstract}{\begin{quotation}  }{\end{quotation}}
\newenvironment{Presented}{\begin{quotation} \begin{center} 
             PRESENTED AT\end{center}\bigskip 
      \centering\begin{large}}{\end{large} \end{quotation}}
\begin{document}
\begin{titlepage}
 \pubblock
\vfill
\Title{Analytic Solution for the Revised Helicity Evolution at Small $x$ and Large $N_c\,$: New Resummed Gluon-Gluon Polarized Anomalous Dimension and Intercept}
\vfill
\Author{Jeremy Borden}
    \Address{Department of Physics, The Ohio State University, Columbus, OH 43210, USA}
    \centering\href{mailto:\Authoremail}{\Authoremail}
%\author{Jeremy Borden}
%         \email[Email: ]{borden.75@buckeyemail.osu.edu}
%	\affiliation{Department of Physics, The Ohio State University, Columbus, OH 43210, USA}
\vfill
\begin{Abstract}
We construct an exact analytic solution of the revised small-$x$ helicity evolution equations derived in \cite{Cougoulic:2022gbk} based on the earlier work \cite{Kovchegov:2015pbl, Kovchegov:2018znm}. The equations we solve are obtained in the large-$N_c$ limit (with $N_c$ the number of quark colors) and are double-logarithmic (summing powers of $\alpha_s \ln^2(1/x)$ with $\as$ the strong coupling constant and $x$ the Bjorken $x$ variable). Our solution provides small-$x$, large-$N_c$ expressions for the flavor-singlet quark and gluon helicity parton distribution functions (PDFs) and for the $g_1$ structure function, with their leading small-$x$ asymptotics given by 
\begin{align}
        \Delta \Sigma (x, Q^2) \sim \Delta G (x, Q^2) 
    \sim g_1 (x, Q^2) \sim \left( \frac{1}{x} \right)^{\alpha_h} ,  \notag
\end{align}
where the exact analytic expression we obtain for the intercept $\alpha_h$ can be approximated by $\alpha_h = 3.66074 \, \sqrt{\frac{\alpha_s \, N_c}{2 \pi}}$. Our solution also yields an all-order (in $\alpha_s$) resummed small-$x$ anomalous dimension $\Delta \gamma_{GG} (\omega)$ which agrees with all the existing fixed-order calculations (to three loops). Notably, our anomalous dimension is different from that obtained in the infrared evolution equation framework developed earlier by Bartels, Ermolaev, and Ryskin (BER) \cite{Bartels:1996wc}, with the disagreement starting at four loops.  Despite the previously reported agreement at two decimal points based on the numerical solution of the same equations \cite{Cougoulic:2022gbk}, the intercept of our large-$N_c$ helicity evolution and that of BER disagree beyond that precision, with the BER intercept at large $N_c$ given by a different analytic expression from ours with the numerical value of $\alpha_h^{BER} = 3.66394 \, \sqrt{\frac{\alpha_s \, N_c}{2 \pi}}$. We speculate on the origin of this disagreement.  
\end{Abstract}
\vfill
\begin{Presented}
DIS2023: XXX International Workshop on Deep-Inelastic Scattering and
Related Subjects, \\
Michigan State University, USA, 27-31 March 2023 \\
\end{Presented}
\vfill
\end{titlepage}

\section{Introduction}
This contribution to the DIS2023 proceedings is based on \cite{Borden:2023tya}. A crucial piece of the proton spin puzzle is the proton's spin content at small Bjorken $x$. Since any experiment can only ever reach some nonzero $x_{\text{min}}$, a theoretical handle on the small-$x$ regime is necessary in order to confidently extrapolate below such an $x_{\text{min}}$. It is in this context that a set of novel small-$x$ helicity evolution equations were derived \cite{Kovchegov:2015pbl, Kovchegov:2016zex, Kovchegov:2017lsr, Kovchegov:2018znm}. An important correction to this KPS (Kovchegov, Pitonyak, and Sievert) evolution was found in \cite{Cougoulic:2022gbk}, resulting in the updated KPS-CTT (`CTT' = Cougoulic, Tarasov, Tawabutr) evolution. The equations are double-logarithmic (resumming powers of $\as \ln^2(1/x)$) and describe the small-$x$ evolution of the `polarized dipole amplitudes,' quantities which ultimately determine the helicity-dependent parton distribution functions (hPDFs) along with the $g_1$ structure function. The KPS-CTT evolution yields an infinite hierarchy of equations, much like that in the unpolarized case \cite{Balitsky:1995ub,Balitsky:1998ya,Kovchegov:1999yj,Kovchegov:1999ua,Jalilian-Marian:1997dw,Jalilian-Marian:1997gr,Weigert:2000gi,Iancu:2001ad,Iancu:2000hn,Ferreiro:2001qy}, but upon taking the large-$N_c$ (or large-$N_c\&N_f$) limit, the infinite hierarchy reduces to a closed system of integral equations.

The updated small-$x$, large-$N_c$ evolution equations were solved numerically in the same paper where they were published \cite{Cougoulic:2022gbk}, yielding small-$x$ asymptotics for the helicity-dependent parton distribution functions and for $g_1$ of
\begin{align}
    \label{num_sol}
    \Delta \Sigma (x, Q^2) \sim \Delta G (x, Q^2) 
    \sim g_1 (x, Q^2) \sim \left(\frac{1}{x}\right)^{3.66\sqrt{\bas}} \hspace{.25cm} , \hspace{.5cm} \text{with} \hspace{.25cm} \bas = \frac{\as N_c}{2\pi}.
\end{align}
The intercept of $3.66\sqrt{\bas}$ appeared to agree with that derived earlier by Bartels, Ermolaev, and Ryskin (BER) using an infrared evolution equations (IREE) approach \cite{Bartels:1996wc}. In addition, an iterative solution of the large-$N_c$ KPS-CTT equations in \cite{Cougoulic:2022gbk} indicated full agreement with the small-$x$, large-$N_c$ part of the glue-glue polarized anomalous dimension $\Delta \gamma_{GG}(\omega)$ out to the existing three-loop order of the fixed-order calculations \cite{Altarelli:1977zs,Dokshitzer:1977sg,Mertig:1995ny,Moch:2014sna} (see also \cite{Zijlstra:1993sh,Moch:1999eb,vanNeerven:2000uj,Vermaseren:2005qc,Blumlein:2021ryt,Blumlein:2021lmf,Davies:2022ofz,Blumlein:2022gpp,Blumlein:2021enk}). Despite this good agreement, an analytic solution of these equations would nevertheless be valuable.

In this work, we have constructed such an analytic solution to the large-$N_c$ KPS-CTT evolution, from which we obtain analytic expressions for the small-$x$ intercept and for an all-order (in $\as$) resummed small-$x$ anomalous dimension $\Delta\gamma_{GG}(\omega)$. These analytic expressions allow us to compare more precisely to the predictions of BER and ultimately reveal a small disagreement in the intercept (at the third decimal point) and in the anomalous dimension (at four loops).

\section{Large-\texorpdfstring{$N_c$}{Nc} Equations}

The large-$N_c$ KPS-CTT evolution equations are written for the (impact-parameter-integrated) polarized dipole amplitudes $G(\xoz^2, zs)$ and $G_2(\xoz^2, zs)$. As defined in \cite{Cougoulic:2022gbk} these amplitudes correspond to sub-eikonal operators inserted between light-cone Wilson lines and are functions of the transverse size squared of the dipole $x_{ij}^2 = |{\un x}_{ij}|^2$ (for $i,j = 0,1,2, \ldots$ labeling the partons and with ${\un x}_{ij} = {\un x}_i - {\un x}_j$ for the two-dimensional transverse vectors ${\un x} = (x^1, x^2)$ in coordinate space) along with the center of mass energy squared $s$ between the original projectile and the target multiplied by the smallest longitudinal momentum fraction $z$ among the two partons making up the dipole. The evolution of the amplitudes $G(\xoz^2, zs)$ and $G_2(\xoz^2, zs)$ also mixes with two additional `neighbor dipole amplitudes' $\Gamma(\xoz^2,\xto^2,zs)$ and $\Gamma_2(\xoz^2,\xto^2,zs)$. These neighbors have the same operator definitions as $G(\xoz^2, zs)$ and $G_2(\xoz^2, zs)$, respectively, but differ in the light-cone lifetime ordering which for the neighbors depends on the adjacent dipole size $\xto^2$ \cite{Kovchegov:2015pbl, Kovchegov:2018znm, Cougoulic:2022gbk}. At large-$N_c$ we have a closed system of four integral equations:

\begingroup
\allowdisplaybreaks
\begin{subequations}\label{largeNc_eqns_unscaled}
\begin{align}
    \label{evolG_unscaled}
    & G(\xoz^2,zs) = G^{(0)}(\xoz^2,zs) + \frac{\as N_c}{2\pi}\int\limits_{\tfrac{1}{s\xoz^2}}^{z}\frac{\mathrm{d}z'}{z'}\int\limits_{\tfrac{1}{z's}}^{\xoz^2}\frac{\mathrm{d}\xto^2}{\xto^2}\Bigg[\Gamma(\xoz^2,\xto^2,z's) + 3 \, G(\xto^2,z's) \\
    & \hspace*{8cm} + 2 \, G_2(\xto^2,z's) + 2 \, \Gamma_2(\xoz^2,\xto^2,z's)\Bigg], \notag \\
    \label{evolGamma_unscaled}
    & \Gamma(\xoz^2,\xto^2,z's) = G^{(0)}(\xoz^2,z's) + \frac{\as N_c}{2\pi}\int\limits_{\tfrac{1}{s\xoz^2}}^{z'}\frac{\mathrm{d}z''}{z''}\int\limits_{\tfrac{1}{z''s}}^{\min\left[\xoz^2,\xto^2\tfrac{z'}{z''} \right] }\frac{\mathrm{d}\xtt^2}{\xtt^2}\Bigg[\Gamma(\xoz^2,\xtt^2,z''s) + 3 \, G(\xtt^2,z''s) \notag \\
    & \hspace*{8cm} + 2 \, G_2(\xtt^2,z''s) + 2 \, \Gamma_2(\xoz^2,\xtt^2,z''s)\Bigg],  \\
    \label{evolG2_unscaled}
    & G_2(\xoz^2,zs) = G_2^{(0)}(\xoz^2,zs) + \frac{\as N_c}{\pi}\int\limits_{\tfrac{\Lambda^2}{s}}^{z}\frac{\mathrm{d}z'}{z'}\int\limits_{\max\left[\xoz^2,\tfrac{1}{z's}\right]}^{\min\left[\tfrac{z}{z'}\xoz^2, \tfrac{1}{\Lambda^2}\right]}\frac{\mathrm{d}\xto^2}{\xto^2}\left[G(\xto^2,z's) + 2 \, G_2(\xto^2,z's) \right], \\
    \label{evolGamma2_unscaled}
    & \Gamma_2(\xoz^2,\xto^2,z's) =  G_2^{(0)}(\xoz^2,z's)  + \frac{\as N_c}{\pi}\int\limits_{\tfrac{\Lambda^2}{s}}^{z'\tfrac{\xto^2}{\xoz^2}}\frac{\mathrm{d}z''}{z''}\int\limits_{\max\left[\xoz^2,\tfrac{1}{z''s}\right]}^{\min\left[\tfrac{z'}{z''}\xto^2, \tfrac{1}{\Lambda^2}\right]}\frac{\mathrm{d}\xtt^2}{\xtt^2}\left[G(\xtt^2,z''s) + 2 \, G_2(\xtt^2,z''s) \right], 
\end{align}
\end{subequations}
\endgroup
where $\Gamma(\xoz^2,\xto^2,z's)$ and $\Gamma_2(\xoz^2,\xto^2,z's)$ are only defined for $\xoz\geq\xto$ and $\Lambda$ is an infrared (IR) cutoff such that we require all the dipole sizes to be $x_{ij} <1/\Lambda$.

Also derived in \cite{Cougoulic:2022gbk} are the following equations which relate the polarized dipole amplitudes to the (dipole) gluon and (flavor-singlet) quark helicity TMDs $g^{G \, dip}_{1L}(x,k_T^2)$ and $g^{S}_{1L}(x,k_T^2)$, along with the hPDFs $\Delta G(x,Q^2)$ and $\Delta \Sigma(x,Q^2)$ and the $g_1$ structure function. $Q(\xoz^2,zs)$ is an additional polarized dipole amplitude, but at large-$N_c$ $Q(\xoz^2,zs) \approx G(\xoz^2,zs)$.

\begin{subequations}\label{distributions+g1}
\begin{align}
    \label{gluon_TMD}
    & g^{G\,dip}_{1L}(x,k_T^2) = \frac{N_c}{\as 2\pi^4}\int \mathrm{d}^2 \xoz\, e^{-i\underline{k}\cdot\underline{x}_{10}} \left[1+\xoz^2\frac{\partial}{\partial\xoz^2}\right]G_2\left(\xoz^2,zs=\frac{Q^2}{x}\right),\\
    \label{quark_TMD}
    & g^{S}_{1L}(x,k_T^2) = \frac{8iN_cN_f}{(2\pi)^5}\int\limits_{\Lambda^2/s}^{1}\frac{\mathrm{d}z}{z}\int \mathrm{d}^2 \xoz\, e^{i\underline{k}\cdot\underline{x}_{10}} \, \frac{\underline{x}_{10}}{\xoz^2}\cdot \frac{\underline{k}}{\underline{k}^2}\left[Q(\xoz^2,zs) + 2 \, G_2(\xoz^2,zs)\right],\\
    \label{gluon_PDF}
    & \Delta G(x,Q^2) = \frac{2 N_c}{\as \pi^2}\left[\left(1+\xoz^2\frac{\partial}{\partial \xoz^2} \right) G_2\left(\xoz^2, zs=\frac{Q^2}{x} \right) \right]_{\xoz^2 = \tfrac{1}{Q^2}}, \\
    \label{quark_PDF}
    & \Delta \Sigma(x,Q^2) = -\frac{N_cN_f}{2\pi^3}\int\limits_{\Lambda^2/s}^{1}\frac{\mathrm{d}z}{z}\int\limits_{\tfrac{1}{zs}}^{\min\left\{\tfrac{1}{zQ^2},\tfrac{1}{\Lambda^2} \right\}} \frac{\mathrm{d}\xoz^2}{\xoz^2}\left[Q(\xoz^2,zs) + 2 \, G_2(\xoz^2,zs)\right],\\
    \label{g1}
    & g_1(x,Q^2) = -\sum_{f}\frac{N_c Z_f^2}{4\pi^3}\int\limits_{\Lambda^2/s}^{1}\frac{\mathrm{d}z}{z} \int\limits_{\tfrac{1}{zs}}^{\min\left\{\tfrac{1}{zQ^2}, \tfrac{1}{\Lambda^2} \right\}} \frac{\mathrm{d}\xoz^2}{\xoz^2}\left[Q(\xoz^2,zs) + 2 \, G_2(\xoz^2,zs) \right].
\end{align}
\end{subequations}
Here ${\un k} = (k^1, k^2)$ is the transverse momentum vector with $k_T = |{\un k}|$, $Z_f$ is the fractional electric charge of the quark, $N_f$ is the number of quark flavors, and Eq.~\eqref{quark_PDF} assumes all quark flavors contribute equally. Then with analytic solutions for the polarized dipole amplitudes $G(\xoz^2,zs)$ and $G_2(\xoz^2,zs)$ (at small-$x$ and large-$N_c$), one can obtain analytic expressions in that same regime for all the quantities in Eqs.~\eqref{distributions+g1}.

\section{Solution}

To solve Eqs.~\eqref{largeNc_eqns_unscaled} analytically we begin by writing the polarized dipole amplitudes $G(\xoz^2,zs)$ and $G_2(\xoz^2,zs)$ as double inverse Laplace transforms in the variables $\ln\left(zs\xoz^2 \right)$ and $\ln\left(1/\xoz^2\Lambda^2\right)$,

\begin{align}
    &G(\xoz^2,zs) = \wint \gint \, e^{\omega\ln(zs\xoz^2) + \gamma \ln \left(\tfrac{1}{\xoz^2\Lambda^2}\right) } G_{\omega\gamma}\,, \\
    &G_2(\xoz^2,zs) = \wint \gint \, e^{\omega\ln(zs\xoz^2) +\gamma \ln \left(\tfrac{1}{\xoz^2\Lambda^2} \right)} \, G_{2\omega\gamma}\,, 
\end{align}

and similarly for their initial conditions/inhomogeneous terms,

\begin{align}
    \label{G0Laplace}
    &G^{(0)}(\xoz^2,zs) = \wint \gint \, e^{\omega\ln(zs\xoz^2) + \gamma \ln \left(\tfrac{1}{\xoz^2\Lambda^2}\right) } G^{(0)}_{\omega\gamma}\,, \\
    \label{G20Laplace}
    &G^{(0)}_2(\xoz^2,zs) = \wint \gint \, e^{\omega\ln(zs\xoz^2) +\gamma \ln \left(\tfrac{1}{\xoz^2\Lambda^2} \right)} \, G^{(0)}_{2\omega\gamma}\,.
\end{align}

Starting from these expressions, one can manipulate the large-$N_c$ evolution equations (Eqs.~\eqref{largeNc_eqns_unscaled}) to obtain expressions for the neighbor dipole amplitudes $\Gamma$, $\Gamma_2$ and constrain the unknown double-Laplace images $G_{\omega\gamma}$, $G_{2\omega\gamma}$. The details of the calculation are omitted here. For the sake of brevity, we just report the most relevant pieces of the solution (omitting the expressions for the neighbor dipole amplitudes $\Gamma$ and $\Gamma_2$ since these do not enter any of quantities in Eqs.~\eqref{distributions+g1}). With that caveat, our solution is

\begin{align}
    &G_2(\xoz^2,zs) = \wint \gint \, e^{\omega\ln(zs\xoz^2) +\gamma \ln \left(\tfrac{1}{\xoz^2\Lambda^2} \right)} \, G_{2\omega\gamma}\,, \\
    &G(\xoz^2,zs) = \wint \gint \, e^{\omega\ln(zs\xoz^2) + \gamma \ln \left(\tfrac{1}{\xoz^2\Lambda^2}\right) } \left[\frac{\omega\gamma}{2 \, \bas}\left(G_{2\omega\gamma} -G^{(0)}_{2\omega\gamma} \right) - 2 \, G_{2\omega\gamma} \right]\,,
\end{align}
with
\begin{align}
    &\gtwg = \gtwg^{(0)} + \frac{\bas}{\omega\left(\gamma-\gamma^-_\omega\right)\left(\gamma-\gamma^+_\omega\right)} \Bigg[2 \left(\gamma-\dw^+\right)\left(G^{(0)}_{\dw^+\gamma} + 2 \, G^{(0)}_{2 \, \dw^+\gamma} \right) \\
    &\hspace{5.5cm}- 2\left(\gamma^+_\omega-\dw^+\right)\left(G^{(0)}_{\dw^+\gamma^+_\omega} + 2 \, G^{(0)}_{2 \, \dw^+\gamma^+_\omega} \right) + 8 \, \dw^-\left(\gtwg^{(0)} - G^{(0)}_{2\omega\gamma^+_\omega} \right) \Bigg]\,, \notag \\
    &\dw^\pm = \frac{\omega}{2}\left[1\pm\sqrt{1-\frac{4\,\bas}{\omega^2}} \right], \\
    \label{gammaomegapm}
    &\gamma^{\pm}_\omega = \frac{\omega}{2}\left[1 \pm \sqrt{1 - \frac{16\,\bas}{\omega^2} \, \sqrt{1-\frac{4\,\bas}{\omega^2}}}  \right]\,,
\end{align}
and with the double-Laplace images of the initial conditions $G^{(0)}_{\omega\gamma}$ and $G^{(0)}_{2\omega\gamma}$ as defined in Eqs.~\eqref{G0Laplace} and \eqref{G20Laplace}.

With these analytic results, one can then use Eqs.~\eqref{distributions+g1} to write down analytic expressions (at small-$x$ and large-$N_c$) for the helicity PDFs and TMDs along with $g_1$. These are

\begin{align}
    \label{gluontmd}
    &g^{G\,dip}_{1L}(x,k_T^2) = \frac{2 \, N_c}{\as \, \pi^3} \, \frac{1}{k_T^2} \wint \gint \, e^{\omega\ln\left(\tfrac{Q^2}{xk_T^2}\right) + \gamma\ln\left(\frac{k_T^2}{\Lambda^2}\right)} \, 2^{2\omega-2\gamma} \, \frac{\Gamma\left(\omega-\gamma+1\right)}{\Gamma\left(\gamma-\omega\right)} \, \gtwg, \\
    \label{gluonhpdf}
    &\Delta G(x,Q^2) = \frac{2 \, N_c}{\as \pi^2} \, \wint \gint \, e^{\omega\ln\left(\tfrac{1}{x}\right) + \gamma\ln\left(\tfrac{Q^2}{\Lambda^2}\right)}\gtwg\,, \\
    \label{quarktmd}
    &g^{S}_{1L}(x,k_T^2) = - \frac{N_f}{\as \, 2 \pi^3} \, \frac{1}{k_T^2}\wint\gint \, \left[  e^{\omega\ln\left(\tfrac{Q^2}{x \, k_T^2}\right) + \gamma\ln\left(\tfrac{k_T^2}{\Lambda^2}\right)} - e^{(\gamma - \omega) \, \ln \left( \tfrac{k_T^2}{\Lambda^2} \right)} \right]\,2^{2\omega-2\gamma}\,\frac{\Gamma\left(1+\omega-\gamma\right)}{\Gamma\left(1-\omega+\gamma\right)} \\ 
    & \hspace*{12cm} \times \,\gamma\left(\gtwg - \gtwg^{(0)}\right)\,, \notag \\
    &\Delta \Sigma(x,Q^2) = - \frac{N_f}{\as \, 2 \pi^2} \, \wint\gint \, \frac{\omega}{\omega-\gamma}\left(\gtwg - \gtwg^{(0)} \right) \, e^{\omega\ln\left(\tfrac{1}{x}\right)} \, e^{\gamma\ln\left(\tfrac{Q^2}{\Lambda^2}\right)} \,, \\
    & g_1(x,Q^2)  = - \frac{1}{2} \sum_f Z_f^2 \, \frac{1}{\as \, 2 \pi^2} \, \wint\gint \, \frac{\omega}{\omega-\gamma}\left(\gtwg - \gtwg^{(0)} \right) \, e^{\omega\ln\left(\tfrac{1}{x}\right)} \, e^{\gamma\ln\left(\tfrac{Q^2}{\Lambda^2}\right)} \,.
\end{align}
Note that the $\Gamma$ appearing in Eqs.~\eqref{gluontmd} and \eqref{quarktmd} is the $\Gamma$-function and not the neighbor dipole amplitude.

\section{Intercept, Anomalous Dimension, and Comparison to BER}
The small-$x$ asymptotics of our solution are governed by the rightmost singularity in the complex-$\omega$ plane. This comes from a branch point in the function $\gamma^-_{\omega}$, defined in Eq.~\eqref{gammaomegapm}. This branch point is

\begin{align}\label{ourintercept}
    \omega = \alpha_h \equiv \frac{4}{3^{1/3}} \, \sqrt{\textrm{Re} \left[ \left( - 9 + i \, \sqrt{111} \right)^{1/3} \right] } \,\sqrt{\frac{\as \, N_c}{2 \pi}} \approx 3.66074 \, \sqrt{\frac{\as \, N_c}{2 \pi}}\,,
\end{align}
and so we have the following small-$x$ asymptotics:
\begin{align}\label{asympt_all}
    \Delta \Sigma (x, Q^2) \sim \Delta G (x, Q^2) 
    \sim g_1 (x, Q^2) \sim g^{G\,dip}_{1L}(x,k_T^2) \sim g^{S}_{1L}(x,k_T^2) \sim \left( \frac{1}{x} \right)^{\alpha_h}\,.
\end{align}

Next, fixing the simple initial conditions $G^{(0)}_2(\xoz^2,zs) = 1 $ and $G^{(0)}(\xoz^2,zs) = 0$, the gluon hPDF in Eq.~\eqref{gluonhpdf} becomes
\begin{align}
    \Delta G(x,Q^2) = \frac{2N_c}{\as \pi^2}\wint e^{\omega\ln\left(\tfrac{1}{x}\right) + \gamma^-_{\omega}\ln\left(\tfrac{Q^2}{\Lambda^2}\right)}\frac{1}{\omega}\,,
\end{align}
from which we can see that our prediction for the resummed all-order in $\as$  $GG$ polarized anomalous dimension at small $x$ and large-$N_c$ (whose subsequent expansion in powers of $\as$ we also show) is
\begin{align}\label{ouranomalousdim}
    \Delta \gamma_{GG}(\omega) = \gamma^-_{\omega} = \frac{\omega}{2}\left[1 - \sqrt{1 - \frac{16\,\bas}{\omega^2}\sqrt{1-\frac{4\,\bas}{\omega^2}} } \ \right] = \frac{4\,\bas}{\omega} + \frac{8\,\bas^2}{\omega^3} + \frac{56\,\bas^3}{\omega^5} + \frac{496\,\bas^4}{\omega^7} + \mathcal{O}(\as^5). 
\end{align}
Reassuringly, this prediction for the anomalous dimension agrees with the fixed-order calculations out to the existing three-loop level (see \cite{Altarelli:1977zs,Dokshitzer:1977sg,Mertig:1995ny,Moch:2014sna,Zijlstra:1993sh,Moch:1999eb,vanNeerven:2000uj,Vermaseren:2005qc,Blumlein:2021ryt,Blumlein:2021lmf,Davies:2022ofz,Blumlein:2022gpp,Blumlein:2021enk}). 

These quantities --- the intercept in Eq.~\eqref{ourintercept} and anomalous dimension in Eq.~ \eqref{ouranomalousdim} --- should be compared to those predicted within the BER IREE formalism \cite{Bartels:1996wc}:

\begin{align}\label{BERintercept}
    \alpha_h^{BER} \equiv \sqrt{\frac{17 + \sqrt{97}}{2}} \, \sqrt{\frac{\as \, N_c}{2 \pi}} \approx 3.66394 \, \sqrt{\frac{\as \, N_c}{2 \pi}},
\end{align}

\begin{align}\label{BERanomalousdim}
     \Delta \gamma^{BER}_{GG} (\omega) = \frac{\omega}{2} \, \left[ 1 - \sqrt{1 - \frac{16 \, \bas}{\omega^2} \, \frac{1 - \frac{3 \, \bas}{\omega^2}}{1 - \frac{\bas}{\omega^2} } } \ \right] = \frac{4 \, \bas}{\omega} + \frac{8 \, \bas^2}{\omega^3} + \frac{56 \, \bas^3}{\omega^5} + \frac{504 \, \bas^4}{\omega^7} + {\cal O} (\as^5) .
\end{align}

While the numerical prefactor of the KPS-CTT intercept in Eq.~\eqref{ourintercept} does indeed round to $3.66$ like the BER intercept in Eq.~\eqref{BERintercept} (as was found in the numerical solution of the large-$N_c$ KPS-CTT equations \cite{Cougoulic:2022gbk}), the two numbers in Eqs.~\eqref{ourintercept} and \eqref{BERintercept} disagree beyond that precision. A similarly small disagreement persists in the anomalous dimensions (comparing that of KPS-CTT in Eq.~\eqref{ouranomalousdim} with that of BER in Eq. \eqref{BERanomalousdim}), with the disagreement only beginning at the four-loop level. At this point, we cannot decisively resolve this (very minor) discrepancy between the predictions of KPS-CTT and BER. However some exploration of the role of hard non-ladder gluons (see Appendix B of \cite{Kovchegov:2016zex}) has provided a possible, though not certain, reason for the differences. In \cite{Bartels:1996wc} BER appear to claim that hard non-ladder gluons should not contribute at DLA, but it does appear possible to construct examples of diagrams which contain hard non-ladder gluons and are DLA included in KPS-CTT evolution (for more details see the appendix of \cite{Borden:2023tya}). In light of the author's limited understanding of the IREE, all of this should be taken as preliminary to be explored further.

\section{Summary}
In this work, we have constructed an analytic solution to the small-$x$ large-$N_c$ KPS-CTT evolution. This solution provides analytic small-$x$, large-$N_c$ expressions for the helicity TMDs, PDFs, and the $g_1$ structure function. Importantly it also provides an analytic expression for the small-$x$ intercept of this evolution and a prediction for the resummed GG polarized anomalous dimension. The expressions for these last two quantities have revealed slight disagreements between the predictions of the KPS-CTT evolution and the BER IREE. While these small discrepancies are certainly interesting and merit further investigation, we believe that at this point the very close agreement between the two different formalisms is promising and should inspire a reasonable degree of confidence in phenomenological applications of the KPS-CTT evolution.

\section{Acknowledgements}
For collaboration, useful discussions, and feedback the author thanks Yuri Kovchegov, Josh Tawabutr, Daniel Adamiak,  Ming Li, Brandon Manley, and Brian Sun. This material is based upon work supported by the U.S. Department of Energy, Office of Science, Office of Nuclear Physics under Award Number DE-SC0004286 and within the framework of the Saturated Glue (SURGE) Topical Theory Collaboration.

%\bibliographystyle{JHEP}
%\bibliography{references}

\providecommand{\href}[2]{#2}\begingroup\raggedright\endgroup

\end{document}